# Three Dimensional Superconductivity in FeSe with $T_c^{zero}$ Up to 10.9 K Induced by Internal Strain

Junyi Ge, Shixun Cao[1], Shujuan Yuan, Baojuan Kang, Jincang Zhang

*Department of Physics, Shanghai University, Shanghai 200444, China*

## ABSTRACT

Polycrystalline sample FeSe was synthesized by a self-flux solution method which shows a zero resistance temperature up to 10.9 K and a $T_c^{onset}$ (90% $\rho_n$, $\rho_n$: normal state resistivity) up to 13.3 K. The decrease of superconducting transition temperature by heat treatment indicates that internal crystallographic strain which plays the same effect as external pressure is the origin of its high $T_c$. The fluctuation conductivity was studied which could be well described by 3D Aslamazov-Larkin (AL) power law. The estimated value of coherence length $\xi_c$=9.2 Å is larger than the distance between conducting layers (~6.0 Å), indicating the three-dimensional nature of superconductivity in this compound.

---

[1] Corresponding author. E-mail: sxcao@shu.edu.cn.

## 1. Introduction

The study of iron-based superconductors has become one of the most attractive fields in condensed matter physics since 2008. Until now more than five families of iron-based superconductors have been reported [1-5] and the $T_c$ has been raised up to 55 K with substitution [1, 6]. Among all of them, the discovery of superconductivity in iron chalcogenide which was called 11-familiy only containing the FeTe(Se) layers is a big surprise. Not like other arsenide superconductors, FeSe itself is superconducting without doping. With increasing in pressure, $T_c$ could be raised from 9 K to 37 K [7-9]. This made the 11-family compounds an idea system in studying superconducting mechanism as well as searching for novel superconductors with potential higher critical temperatures. In the study of 1111-phase FeAs superconductors, one common used method to improve the critical temperature is by hole or electron doping, which could increase the carrier concentrations, *e.g.* in $SmFeAsO_{1-x}F_x$ [10] and $CeFeAsO_{1-x}F_x$ [11]. However, in iron-based chalcogenides, the substitution effect on FeSe system both in Fe and Se site seems negligible, even negative to its $T_c$. *e.g.* in $FeSe_{1-x}S_x$, $T_c$ was slightly increased with the maximum value of $T_c^{zero}$=8.4 K, while in $Fe_{1-x}Co_xSe$ and $Fe_{1-x}Ni_xS$ $T_c$ was suppressed with doping[12]. Until now, the highest $T_c$ was obtained in $FeSe1_{-x}Te_x$ system with $T_c$ onset about 15 K [13], much smaller than $T_c^{max}$ of FeSe under pressure. This is very unintelligible since doping is always thought an effective way to introduce chemical pressure into the system which may have the same effect as external hydrostatic pressure.

Here in this paper we report the observation of $T_c^{zero}$ up to 10.9 K in FeSe at ambient pressure. As far as we know, this is the highest $T_c$ among all the FeSe bulk samples without Te in the system reported until now. Through heat treatment and the study of resistivity as well as magnetization measurements, we found that $T_c$ was decreased to normal value (~8.6 K), suggesting the relative high $T_c$ phase in our sample is strongly related to the internal crystallographic strain. Temperature dependence of resistivity under various magnetic fields was also measured. The upper critical field is estimated to be 40.7 T by using WHH formula. The fluctuation conductivity was studied which could be well described by 3D AL power law, indicating the three-dimensional nature of superconductivity in this compound.

## 2. Experimental

Polycrystalline sample with a nominal composition of $Fe_{1.03}Se$ was prepared from powders of Fe (99.9%), Se (99.999%). The mixed powders were thoroughly ground in a mortar under the protection of argon atmosphere, pressed into pellets and then sealed in an evacuated quartz tube. The tube was heated in 700℃ for 20 hours and then cooled with furnace. The reacted sample was reground into powders, loaded in a double quartz tube, heated in an optical floating-zone furnace (FZ-T-10000-H-VI-P-SH) with 4×500 W halogen lamps installed as infrared radiation sources. The tube is rotated at a rate of 20 rpm and moved down at a transition rate of 4 mm/h.

Powder X-ray diffraction measurement was carried out with X-ray diffractometer (D/MAX-2550) using Cu $K_α$ radiation. Electrical resistivity measurements were carried out in the temperature range of 2–300 K and in magnetic fields up to 9 T by the standard four-probe method using a Quantum Design physical-property-measurement system (PPMS-9). Four contact wires were painted onto the samples using silver paste. DC susceptibility was measured using VSM (option attached to the PPMS-9).

## 3. Results and discussion

Fig. 1(a) shows the SEM image of the as-grown sample. It can be seen that the sample has a hole-like surface. The X-ray diffraction pattern of the as-grown sample was measured as shown in fig. 1(b). Most peaks could be well indexed using P4/nmm space group. Yet there are still some impurities indexed by '*'. As the composition of the as-grown crystal always deviates from the nominal composition like been reported in Ref. [14, 15], here the refined composition determined by EDX is FeSe as shown in fig. 1(c).

In fig. 2, we show the temperature dependence of resistivity of FeSe from 2 K to room temperature. The resistivity began to decrease at 14.0 K as can be seen from the top inset of fig. 2, and dropped to zero at $T_c^{zero}$=10.9 K. This is 2-3 K higher than the $T_c^{zero}$ of FeSe reported before both in single crystal or polycrystalline samples [16,17]. Temperature dependence of magnetic susceptibility was measured in order to confirm the superconducting transition. As can be seen in the bottom inset of fig. 2, the magnetic onset of susceptibility appears at 10.9 K which is the same as zero resistance temperature. The volume fraction of superconductivity reaches as high as 14%, suggesting considerable bulk superconductivity in our sample. The relative high $T_c$ superconducting phase is surprising, since there is no charge doping or external pressure in our sample. In order to further improve the superconducting phase, the as-grown sample was heated to 400℃ in vacuum and hold for 10 hours. Usually, the annealing procedure could enhance the superconductivity in 11-family compounds [14, 18]. As can be seen in fig. 3, the diamagnetic signal became more pronounced after annealing. However, to one's surprise, the transition temperature dropped to the value of 8.6 K which is consistent with those reported before [14, 19]. The inset of fig. 3 shows the temperature dependence of resistivity in the as-grown and annealed sample which also confirms the same effect. All these data show that the internal crystallographic strain seems to play the same role as the external pressure, which might be the origin of the relative higher $T_c$ superconducting phase in the as-grown sample. Similar results have also been observed in ref. [20]. In ref. [20], due to the lattice distortion induced by internal strain, $SrFe_2As_2$ became superconducting without any hole or electron doping at ambient pressure. Also, we have noticed that in ref. [21], the authors reported the onset superconducting transition at 24 K in $Fe_{1+x}Se$ related to the decrease of unit cell volume, although the fraction of superconducting phase seems very small and $T_c^{zero}$ is around 6 K. The internal strain may come from the special synthesization method since all the preparations of the samples mentioned above include the process of melting.

The normal-state resistivity $\rho(T)$ shows two different transport behaviors. Below the temperature $T_s$=75 K, $\rho(T)$ follows a strict linear temperature dependence which could be well described by a fit with the form of $\rho(T)=\rho_0+AT$. However, above $T_s$, it can be described by another approximate form of $\rho_0+AT+BT^2$. This behavior is reminiscent of the normal-state resistivity in organic and 122-phase FeAs superconductors [22], in which the linear transport properties were suggested to be related to magnetic fluctuations associated with SDW. Yet, in FeSe system, no magnetic order has been detected by different measurement methods [23, 24], and the abnormally of resistivity at $T_s$ is ascribed to the structure phase transition [4, 25]. Further study is needed to make clear the relationship between structure transition and magnetic order.

Temperature dependence of resistivity under applied magnetic fields was measured as shown in fig. 4. Three characteristic temperatures of the superconducting transition were defined: the onset temperature $T_c^{onset}$ (90 % of the normal state resistivity $\rho_n(H, T)$), the mid-point temperature $T_c^{mid}$ (50 % of $\rho_n(H, T)$), and the zero-resistivity temperature $T_c^{offset}$ (10 % of $\rho_n(H, T)$) according to the definition reported in ref. [26-28]. The values of $T_c$ at 0 T were determined to be $T_c^{onset}(0) = 13.3$ K, $T_c^{mid}(0) = 12.2$ K, and $T_c^{offset}(0) = 11.4$ K, respectively. The upper critical field ($\mu_0 H_{c2}$) was plotted in the inset of Fig. 4. All three curves show almost linear dependence with temperature, and no upturn curvature near 0 T was observed like been reported in FeTe$_{0.75}$Se$_{0.25}$ [28]. The slopes of $\mu_0 H_{c2}$ at $T_c^{onset}(0)$, $T_c^{mid}(0)$, and $T_c^{offset}(0)$ are -4.44 T/K, -3.39 T/K, and -3.25 T/K, respectively. From the conventional Werthamer-Helfand-Hohenberg (WHH) theory with the formula:

$$\mu_0 H_{c2}(0) = -0.69 T_c (d\mu_0 H_{c2} / dT)\big|_{T=T_c} \qquad (1)$$

The estimated upper critical magnetic field at zero temperature are $\mu_0 H_{c2}^{onset}$=40.7 T, $\mu_0 H_{c2}^{mid}$=28.5 T, and $\mu_0 H_{c2}^{offset}$=25.6 T, respectively, relatively smaller compared with FeAs-based superconductors.

In high-$T_c$ superconductors, due to the finite Cooper pair formation the resistivity often shows a rounded curvature deviating from the normal state resistivity in the vicinity of superconducting transition. The experimental conductivity σ was formed by two parts: The normal state conductivity $σ_n$; the fluctuation conductivity Δσ which is also called paraconductivity. The fluctuation conductivity also contains two parts: the Aslamazov-Larkin term [29] $σ_{AL}$ and the Maki-Thompson term [30, 31] $σ_{MT}$. The latter part was always found to be less divergent [32, 33] and hence negligible. Hence, the experimental conductivity could be described as:

$$\sigma = \sigma_n + \Delta\sigma \approx \sigma_n + \sigma_{AL}. \qquad (2)$$

Here the leading contributions to paraconductivity take two forms due to the coupling strength between the conducting planes,

$$\delta\sigma_{AL}^{3D} = \frac{e^2}{32\hbar\xi_c}\frac{1}{\sqrt{\varepsilon}} \qquad (3)$$

$$\delta\sigma_{AL}^{2D} = \frac{e^2}{16\hbar d}\frac{1}{\varepsilon} \qquad (4)$$

where d is the distance between the conducting layers, $\xi_c$ is the coherence length along the direction perpendicular to the layers, ε=log(T/$T_c^{MF}$)≈(T−$T_c^{MF}$)/$T_c^{MF}$, $T_c^{MF}$ is the mean field transition temperature. According to Lawrence and Doniach [34], in layered high temperature superconductors, the paraconductivity mainly come from the contribution of in-plane part which takes the form of eq. (4). When the coupling between the layers becomes strong, the paraconductivity will turn to the three dimensional form (eq. (3)). Thus, this makes the study of fluctuation conductivity an effective way to study the dimensions of superconductivity.

From eq. (2) we can see the choice of normal state conductivity $\sigma_n$ (or resistivity $\rho_n$) is crucial for the confirmation of paraconductivity. In fig. 2, $\rho(T)$ shows two different behaviors up and below $T_s$. As is known, $FeSe_{1-x}$ experiences a structure transition between 70 and 100 K. Yet in Fe-rich compositions, the structure transition is reported to disappear, and superconductivity is not found, suggesting that the structure transition is strongly related to the occurrence of superconducting phase in iron selenides. In another words, the superconducting phase was formed after the structure transition. Therefore, to determine the normal state resistivity, we can only consider the interval between $T_c$ and $T_s$ (precisely, in the region between 25 K and 55 K, where fluctuation is assumed to be negligible). Therefore, the normal state resistivity $\rho_n$ could be described using the fit $\rho_n(T)=a+bT$ with the fitting parameters a=8.99 m$\Omega$ cm, b=0.37 m$\Omega$/K.

In fig. 5 we show the comparison between the experimental paraconductivity (black square) and the 2D (green line) and 3D (yellow line) expressions of AL paraconductivity. One can see the experimental data in a log-log plot could be well described by the 3D power law [eq. (2)] with $\xi_c$=9.2 Å, while the 2D AL results with a parameter d=6.0 Å is about 2 orders of magnitude larger than the experimental data. $T_c^{MF}$ derived from the 3D AL fitting curve is 13.6 K, 0.3 K larger than $T_c^{onset}$, which is reasonable. The SC fluctuations persist up to $\varepsilon = 0.2$ and then drop drastically below the 3D AL fitting curve. Similar behavior has also been observed in $SmFeAsO_{0.8}F_{0.2}$ which is ascribed to the interband coupling mechanism [35]. According to Pallecchi et

al.'s report [36], the F-doped SmFeAsO shows a 2D nature of superconductivity. This is understandable since in iron-based superconductors the FeAs(Se,Te) layers are thought to be crucial for the superconductivity while the 'charge reservoir' layers are not as necessary as that in high-$T_c$ cuprates. So the coupling between the conducting layers of FeSe is much larger than that of 1111-phase superconductors due to its relative small interdistance.

## 4. Conclusions

In conclusion, FeSe with $T_c^{onset}$ up to 13.3 K was synthesized by flux method. The decrease of superconducting transition temperature by heat treatment indicates that internal crystallographic strain which plays the same effect as external pressure is the origin of high $T_c$. The fluctuation conductivity in the vicinity of $T_c$ could be well described by 3D fitting curve. The estimated value of coherence length $\xi_c$=9.2 Å is larger than the interplanar spacing d=6.0 Å, indicating a 3D nature of superconductivity in this compound.

**Acknowledgment**

This work is supported by the National Natural Science Foundation of China (NSFC, No.50932003, 10774097), and the Science and Technology Innovation Fund of the Shanghai Education Committee (No.09ZZ95), and the Science & Technology Committee of Shanghai Municipality (No.08dj1400202, 10ZR1411000, 08521101502).

**Reference**

[1] Kamihara Y, Watanabe T, Hirano M and Hosono H, *J. Am. Chem. Soc.*, **130** (2008) 3296–3297.

[2] Rotter M, Tegel M, Johrendt D, Schellenberg I, HermesWand ottgen R P, *Phys. Rev. B*, **78** (2008) 020503.

[3] Wang W, Liu Q, Y-Lv, Gao W, Yang L X, Yu R C, Li F Y and Jin C, *Solid State Comm.*, **148** (2008) 538–540.

[4] Hsu F C, Luo J Y, Yeh K W, Chen T K, Huang T W, Wu P M, Lee Y C, Huang Y L, Chu Y Y, Yan D C and Wu M K, *Proc. Natl. Acad. Sci U.S.A.*, **105** (2008) 14262.

[5] Ogino H, Matsumura Y, Katsura Y, Ushiyama K, SHorii, Kishio K and Shimoyama J, *Supercond. Sci. Technol.*, **22** (2009) 075008.

[6] Ren Z, Lu W, Yang J, Yi W, Shen X L, Li Z C, Che G C, Dong X L, Sun L L, Zhou F and Zhao Z X, *Chin. Phys. Lett.*, **25** (2008) 2215–2216.

[7] Y. Mizuguchi, F. Tomioka, S. Tsuda, T. Yamaguchi, and Y. Takano, *Appl. Phys. Lett.*, **93** (2008)152505.

[8] S. Margadonna, Y. Takabayashi, Y. Ohishi, Y.Mizuguchi, Y. Takano, T. Kagayama, T. Nakagawa, M. Takata, and K. Prassides, arXiv:0903.2204 (2009).

[9] S. Medvedev, T. M. McQueen, I. A. Troyan, T. Palasyuk, M. I. Eremets, R. J. Cava, S. Naghavi, F. Casper, V. Ksenofontov, G.Wortmann and C. Felser, *Nature Materials*, **8** (2009) 630.

[10]R. H. Liu, G.Wu, T.Wu, D. F. Fang, H. Chen, S.Y. Li, K. Liu, Y. L. Xie, X.

F.Wang, R. L. Yang, L. Ding, C. He, D. L. Feng, and X. H. Chen, *Phys. Rev. Lett.*, **101** (2008) 087001.

[11] Jun Zhao, Q. Huang, Clarina de la Cruz, Shiliang Li, J. W. Lynn, Y. Chen, M. A. Green, G. F. Chen, G. Li, Z. Li, J. L. Luo, N. L. Wang, Pengcheng Dai, *Nature Materials*, **7** (2008) 953-959.

[12] Y. Mizuguchi, F. Tomioka, S. Tsuda, T. Yamaguchi, Y. Takano, *J. Phys. Soc. Jpn.*, **78** (2009) 7.

[13] M. H. Fang, H. M. Pham, B. Qian, T. J. Liu, E. K. Vehstedt, Y. Liu, L. Spinu, Z. Q. Mao, *Phys. Rev. B*, **78** (2008) 224503.

[14] K.W. Yeh, C.T. Ke, T.W. Huang, T.K. Chen, Y.L. Huang, P.M. Wu, M.K. Wu, *Cryst. Growth Des.*, **9 (11)** (2009) 4847.

[15] B. C. Sales, A. S. Sefat, M. A. McGuire, R. Y. Jin, and D. Mandrus, Y. Mozharivskyj, *Phys. Rev. B*, **79** (2009) 094521.

[16] U. Patel,1 J. Hua, S. H. Yu, S. Avci, Z. L. Xiao, H. Claus, J. Schlueter, V. V. Vlasko-Vlasov, U. Welp, and W. K. Kwok, *Appl. Phys. Lett.*, **94** (2009) 082508.

[17] Yoo Jang Song, Jong Beom Hong, Byeong Hun Min, Kyu Jun Lee, Myung Hwa Jung, Jong-Soo Rhyee, Yong Seung Kwon, arXiv:0911.2045.

[18] T. Taen, Y. Tsuchiya, Y. Nakajima, and T. Tamegai, *Phys. Rev. B*, **80** (2009) 092502.

[19] D Braithwaite, B Salce, G Lapertot, F Bourdarot, C Marin, D Aoki and M Hanfland, *J. Phys.: Condens. Matter*, **21** (2009) 232202.

[20] S. R. Saha, N. P. Butch, K. Kirshenbaum, Johnpierre Paglione and P. Y. Zavalij,

*Phys. Rev. Lett.*, **103** (2009) 037005.

[21] A. E. Karkin, A. N. Titov, E. G. Galieva, A. A. Titov, B. N. Goshchitskii, arXiv:0911.1194 (2009).

[22] N. Doiron-Leyraud, P. Auban-Senzier, S. R. d. Cotret, A. Sedeki, C. Bourbonnais, D. Jérome, K. Bechgaard and L. Taillefer, arXiv 0905.0964 (2009).

[23] T. M. McQueen, Q. Huang, V. Ksenofontov, C. Felser, Q. Xu, H. Zandbergen, Y. S. Hor, J. Allred, A. J. Williams, D. Qu, J. Checkelsky, N. P. Ong and R. J. Cava, *Phys. Rev. B*, **79** (2009) 014522.

[24] T. M. McQueen, A. J. Williams, P. W. Stephens, J. Tao, Y. Zhu, V. Ksenofontov, F. Casper, C. Felser, R. J. and R. J. Cava, ArXiv: 0905.1065 (2009).

[25] S. Margadonna, Y. Takabayashi, M. T. McDonald, K. Kasperkiewicz, Y. Mizuguchi, Y. Takano, A. N. Fitch, E. Suard and K. Prassides, *Chemical Communications*, **43** (2008) 5607-5609.

[26] K.W. Yeh, C.T. Ke, T.W. Huang, T.K. Chen, Y.L. Huang, P.M. Wu, M.K. Wu, arXiv:0908.2855 (2009).

[27] F. Hunte, J. Jaroszynski, A. Gurevich, D. C. Larbalestier, R. Jin, A. S. Sefat, M. A. McGuire, B. C. Sales, D. K. Christen, and D. Mandrus, Nature 453 (2008) 903.

[28] Takanori Kida, Takahiro Matsunaga, Masayuki Hagiwara, Yoshikazu Mizuguchi, Yoshihiko Takano, Koichi Kindo, J. Phys. Soc. Jpn. 78 (2009) 113701.

[29] L. G. Aslamazov, A. I. Larkin, *Phys. Rev. Lett.*, **A26** (1986) 238.

[30] K. Maki, *Prog. Theor. Phys.*, **39** (1968) 897; **40** (1968) 193.

[31] R. S. Thompson, *Phys. Rev. B*, **1** (1970) 327.

[32] S. K. Yip, *Phys. Rev. B*, **41** (1990) 2612–2615.

[33] C. Dicastro, C. Castellani, R. Raimondi, A. Varlamov, *Phys. Rev. B*, **42** (1990) 10211.

[34] W. E. Lawrence and S. Doniach, in **Proc. 12$^{th}$ Int. Conf. on Low Temp. Phys. (Kyoto)** (1970).

[35] L. Fanfarillo, L. Benfatto, S. Caprara, C. Castellani, and M. Grilli, *Phys. Rev. B*, **79** (2009) 172508.

[36] I. Pallecchi, C. Fanciulli, M. Tropeano, A. Palenzona, M. Ferretti, A. Malagoli, A. Martinelli, I. Sheikin, M. Putti and C. Ferdeghini, *Phys. Rev. B*, 79 (2009) 104515.

**Figure Captions**

**Fig. 1** Scanning electron microscopic picture (a) and X-ray diffraction patterns (b) of the as-grown sample. The red rectangle marks the position where we took the EDX spectrum. (c) The EDX spectrum taken from one piece of the sample.

**Fig. 2** (Color online) Temperature dependence of resistivity, the arrow shows the temperature of structure phase transition. The left inset shows resistivity and the d$\rho$/dT curve at low temperature regime. The right inset shows the magnetic susceptibility as a function of temperature in a field of 50 Oe. The arrows in both insets show the onset transition temperature.

**Fig. 3** (Color online) Magnetic susceptibility (main panel) and normalized resistivity (inset) as a function of temperature for the as-grown and annealed sample.

**Fig. 4** (Color online) Temperature dependence of the electrical resistivity in dc magnetic fields up to 9 T. The inset displays the temperature dependence of resistive upper critical field $\mu_0 H_{c2}(T)$ at three defined temperatures.

**Fig. 5** (Color online) Fluctuation conductivity $\Delta\sigma$ as a function of $\varepsilon$ in a Log-Log scale. The solid lines represent the 3D (yellow) and 2D (green) Aslamazov-Larkin behavior, respectively. For the 3D paraconductivity a coherence length $\xi_c$=9.2 Å has been used while for the 2D case the structural distance between layers d=6.0 Å has been inserted.

**FIG. 1**

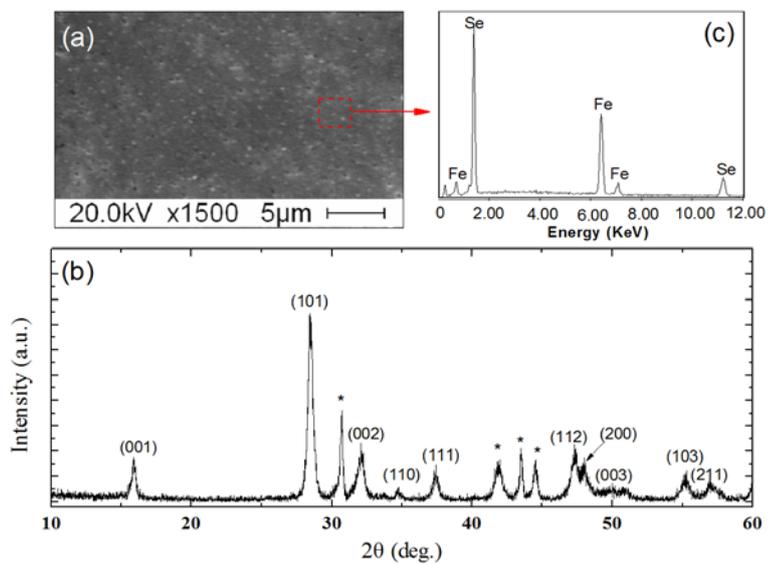

**FIG. 2**

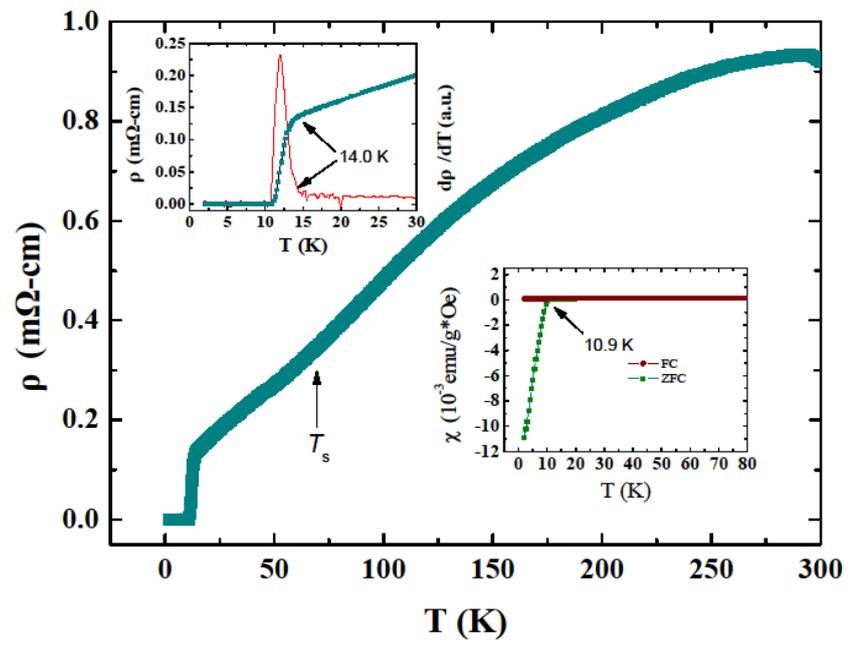

**FIG. 3**

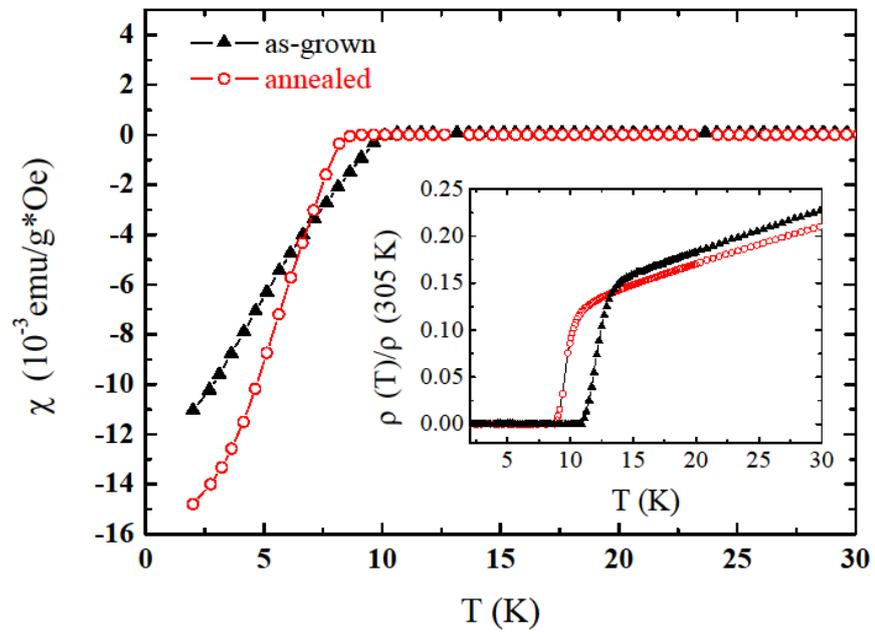

**FIG. 4**

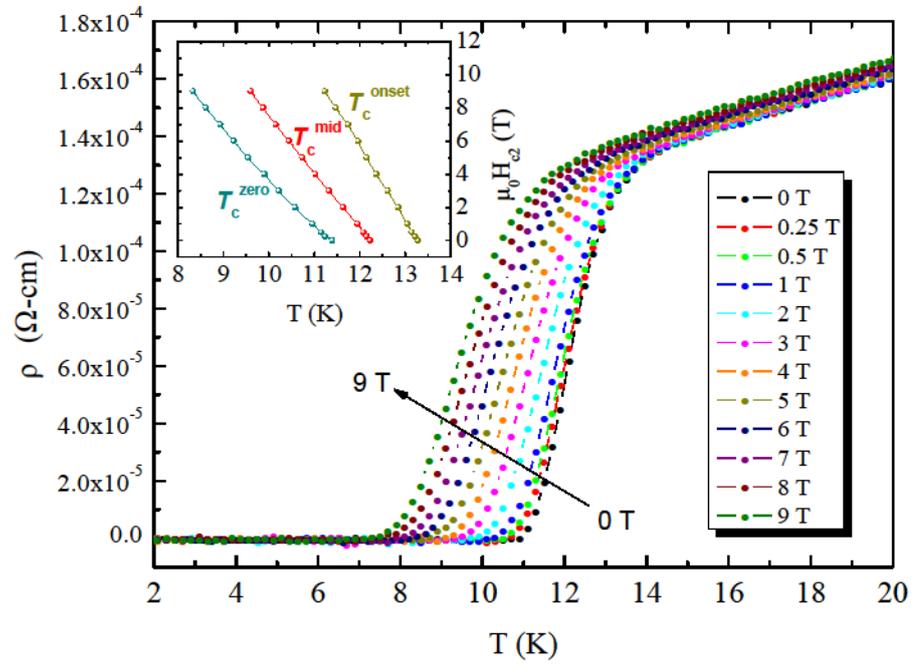

**FIG. 5**

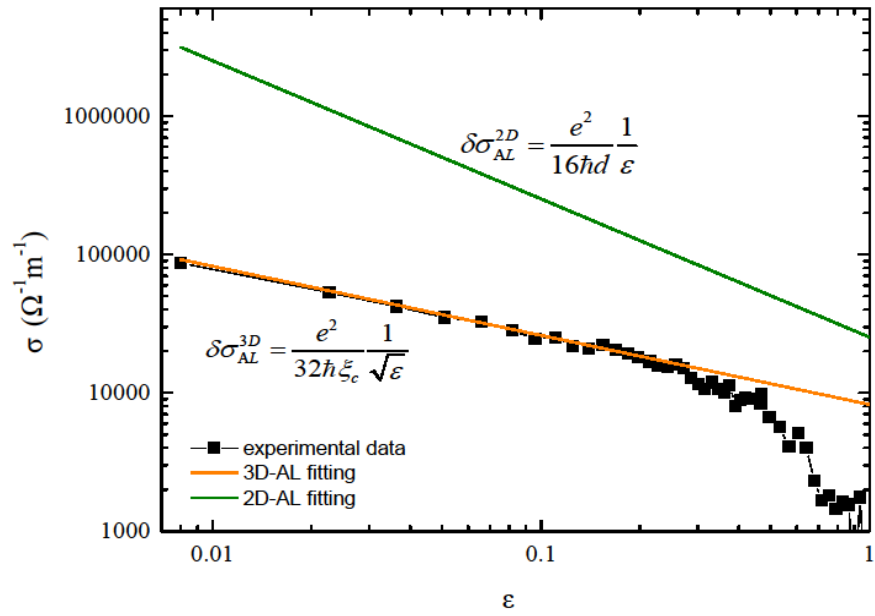